# Predicting Financial Market Trends using Time Series Analysis and Natural Language Processing

Ali Asgarov
Compter Science and Data Analytics
School of Information Technologies
and Engineering
ADA University & George Washington
University
Baku, Azerbaijan
ali.asgarov@gwmail.gwu.edu

*Abstract* — Forecasting financial market trends through time series analysis and natural language processing poses a complex and demanding undertaking, owing to the numerous variables that can influence stock prices. These variables encompass a spectrum of economic and political occurrences, as well as prevailing public attitudes. Recent research has indicated that the expression of public sentiments on social media platforms such as Twitter may have a noteworthy impact on the determination of stock prices. The objective of this study was to assess the viability of Twitter sentiments as a tool for predicting stock prices of major corporations such as Tesla, Apple. Our study has revealed a robust association between the emotions conveyed in tweets and fluctuations in stock prices. Our findings indicate that positivity, negativity, and subjectivity are the primary determinants of fluctuations in stock prices. The data was analyzed utilizing the Long-Short Term Memory neural network (LSTM) model, which is currently recognized as the leading methodology for predicting stock prices by incorporating Twitter sentiments and historical stock prices data. Our analysis yielded findings indicating that Twitter sentiments possess significant potential as an informative resource for forecasting stock prices. The models utilized in our study demonstrated a high degree of reliability and yielded precise outcomes for the designated corporations. In summary, this research emphasizes the significance of incorporating public opinions into the prediction of stock prices. The application of Time Series Analysis and Natural Language Processing methodologies can yield significant scientific findings regarding financial market patterns, thereby facilitating informed decision-making among investors. The results of our study indicate that the utilization of Twitter sentiments can serve as a potent instrument for forecasting stock prices, and ought to be factored in when formulating investment strategies.

## I. Introduction

External factors, such as the influence of social media, can have a significant impact on the value of stocks, thereby contributing to the market's widely acknowledged volatility. The aforementioned phenomenon has resulted in an expanding field of research focused on exploring the feasibility of utilizing Twitter sentiment analysis data for predicting fluctuations in stock prices. Previous research has investigated the correlation between emotions expressed on Twitter and fluctuations in stock prices. According to a research study, the accuracy rate for predicting market and sector changes was found to be 68% [1]. The study conducted by Mao, Wei, Wang, and Benyuan in 2012 involved the correlation of daily Tweets that mentioned S&P 500 with the stock indicators of S&P 500. The sentiment expressed by the general public on social media platforms can impact the market demand for a corporation's shares, subsequently affecting its stock valuation. The positive sentiment expressed through news and tweets may lead to an increase in the stock price of a company [2]. Hence, it is imperative for all stakeholders of the stock market to comprehend and scrutinize the public opinions on Twitter.

The objective of sentiment analysis is to extract the subjective emotions and attitudes conveyed in a written or spoken language. Our research employed BERT (Bidirectional Encoder Representations from Transformers) to derive metrics such as "positive," "negative," and "neutral." During the timeframe of April 2022 to April 2023, our analysis focuses on Twitter posts pertaining to the stock market that mention the companies Apple and Tesla. The heterogeneity of the product and service portfolios of these enterprises enhances the veracity of our prognostic modeling outcomes. When evaluating the impact of Twitter sentiment on stock market fluctuations, these enterprises possess robust financial foundations, rendering the correlation and predictive outcomes valuable for investors, business owners, and the companies themselves.

In addition to the aforementioned, customer insights can help investors design risk mitigation methods, entrepreneurs implement creative marketing for startups, and existing companies create cutting-edge product plans. It is essential in today's ever-changing and uncertain global scene to understand the public's sentiment as it evolves from feelings to opinions. It's important to catch feelings before they harden into beliefs since it's hard to shift people's minds once they've made up their minds. Understanding your customers and the wider public requires constant vigilance over real-time social media channels like Twitter.

To monitor a wide range of emotions and enhance stock price projections for companies, our suggested model uses BERT, a state-of-the-art language model for sentiment analysis. The purpose of this research is to answer three important issues. The primary objective is to study how different Twitter sentiment measurements and the stock price characteristics of major corporations like Apple, and Tesla relate to one another. By digging into this link, we can see how Twitter sentiment affects stock prices for these companies and which sentiment indicators correlate strongly with the stock data.

The secondary aim of this investigation is to determine if there exists a correlation between changes in the overall sentiment metrics obtained from tweets of the preceding day and the response of the stock market. By employing this approach, we can observe the change in time of sentiments expressed on the Twitter platform. The study aims to assess the effectiveness of utilizing the stock prices and Twitter sentiment values from the previous day to forecast the stock prices of a company for the following day.

We collect Twitter data from the preceding day and merge it with stock market information for statistical analysis to evaluate the validity of our hypotheses regarding the correlation between tweets and the stock prices of prominent corporations. We employ the variables of 'positivity,' 'negativity,' and the previous day's stock price as predictors to anticipate the future values of stock prices. Time-series data that is sequential in nature, such as stock prices, exhibit a high degree of dependence on previous values. This article presents the Long Short-Term Memory (LSTM) neural network model. The Long Short-Term Memory (LSTM) variant of Recurrent Neural Network (RNN) is a significant and innovative RNN variant as it has the ability to retain both long- and short-term values while allowing the neural network to store only pertinent information [3].

The present study highlights the utilization of LSTM models for the prediction of stock prices. Additional particulars regarding the methodology can be found in the corresponding section. In the current competitive scenario, it is crucial to utilize sophisticated techniques that combine public sentiment with stock market information to comprehend and forecast forthcoming patterns. The objective of this investigation is to formulate a methodology that can serve as a basis for additional investigation and enhancement by scientists.

## II. BACKGROUND

The financial market is a complex system that is influenced by a wide range of factors. These factors can range from economic and political occurrences to public attitudes and opinions. Understanding the interplay between these various factors is essential for predicting market trends and making informed investment decisions.

In recent years, there has been growing interest in the use of natural language processing (NLP) and sentiment analysis techniques to analyze social media data for predicting stock prices. Social media platforms, such as Twitter, provide a wealth of data that can be used to gauge public sentiment and attitudes towards specific companies or products. This information can then be used to predict how the stock prices of those companies might fluctuate in the future.

However, accurately predicting stock prices using social media data is a challenging undertaking. Stock prices are influenced by a variety of factors, and social media sentiment is just one of many variables that must be taken into account. Additionally, social media data can be noisy and difficult to interpret, making it challenging to extract meaningful insights.

Despite these challenges, recent research has suggested that there is a strong correlation between Twitter sentiment and stock prices. Specifically, studies have found that positive or negative sentiment on Twitter can be a reliable predictor of future stock prices. This has led to increased interest in the use of NLP and sentiment analysis techniques for predicting stock prices.

One of the primary advantages of using NLP and sentiment analysis for predicting stock prices is that it can provide a more holistic view of the market. By analyzing social media data, investors can gain insights into the attitudes and opinions of a wide range of individuals, including both expert analysts and everyday consumers. This can help to identify potential market trends and opportunities that may be missed by traditional analysis methods.

In recent years, deep learning models, such as Long-Short Term Memory (LSTM) neural networks, have emerged as a leading methodology for predicting stock prices using social media data. LSTM networks are designed to analyze time-series data, making them well-suited for analyzing stock prices over time. Additionally, LSTM networks can be trained to analyze the sentiment of social media data, allowing them to extract insights from the noisy and complex data.

In summary, the use of time series analysis and sentiment analysis for predicting stock prices is a rapidly evolving field with significant potential. By incorporating social media data into traditional analysis methods, investors can gain a more comprehensive understanding of market trends and make more informed investment decisions. However, accurately predicting stock prices using social media data remains a challenging undertaking, and additional research is needed to further refine the methodologies and techniques used.

## III. RELATED WORK

In recent years, the analysis of the stock market with a focus on public opinion has incorporated social media data to a greater extent. The upward trend observed in the stock market during recent decades has stimulated a renewed academic interest in forecasting stock prices. In contemporary times, there has been a surge of curiosity regarding the potential contribution of Natural Language Processing (NLP) in the domain of financial forecasting. Tetlock (2007) conducted a study utilizing Wall Street Journal data to examine the correlation between media sentiment and stock market trends [4]. The results indicated that the presence of highly negative sentiment had an adverse impact on the stock market. Mao, Counts, and Bollen (2011) conducted an analysis of Twitter feeds, news headlines, and other internet data sources. They utilized sentiment tracking measures to predict the financial market's value [5].
Research conducted by Shah, Isah, and Zulkernine (2019) suggests that the sentiments can have an impact on short-term market fluctuations [6]. This phenomenon may result in divergences in the market value of a firm's stocks and its fundamental worth. The findings lend support to our hypothesis that the sentiments conveyed on social media platforms such as Twitter could serve as a significant indicator of daily market trends. Bollen et al. (2011) conducted a study on the correlation between the Dow Jones Industrial Average (DJIA) and collective emotional states derived from extensive Twitter feeds. The study analyzed the relationship between the two variables over a period of time [7]. In their study, Bollen employed Granger causality analysis and Self-Organizing Fuzzy Neural Networks to predict changes in DJIA closing values. They achieved a precision of 86.7% in their predictions. To accomplish this, sentiment analysis tools such as 'OpinionFinder' and 'Google Profile of Mood States (GPOMS)' were utilized to examine the correlation between sentiment lag values and stock prices. Our research employs the BERT sentiment analysis tool. Our study aims to analyze the influence of tweet sentiments that are aggregated for the previous day period on current stock prices, instead of using lag values. As per the studies conducted by Pagolu et al. (2016) and Kordonis, sentiment analysis of Twitter can aid in forecasting the stock market's trajectory [8]. The obtained outcomes have confirmed the soundness of our choice to construct prognostic models and have stimulated us to utilize Twitter sentiment analysis as a means to predict stock market prices. Dickinson & Hu (2015) [9] employed the Pearson correlation coefficient to investigate the correlation between fluctuations in stock prices and public opinion. As an alternative, Spearman's rank correlation tests were employed, which are not dependent on the normality assumption of the analyzed data. Namini conducted a comparative analysis between Long Short-Term Memory (LSTM) deep learning algorithms and Autoregressive Integrated Moving Average (ARIMA) conventional techniques for time-series data forecasting, in order to assess the efficacy of the former [10]. Their findings were presented as a contrast between the two methods. According to their empirical results, the Long Short-Term Memory (LSTM) model outperforms the Autoregressive Integrated Moving Average (ARIMA) model, resulting in a mean reduction of errors by 84%. This comprehensive analysis of the relevant scientific literature illustrates the superiority of LSTM models over Regressor and ARIMA models in the realm of stock price prediction. Consequently, supplementary experiments were carried out utilizing LSTM model configurations to evaluate the impact of hyperparameter tuning.

LI Bing conducted sentiment analysis [11] on Twitter data consisting of text-based tweets using Natural Language Processing (NLP) techniques to eliminate uncertain tweet data and depict the general public's sentiment. The authors employed a data-mining methodology to identify patterns in the correlation between public sentiment and actual fluctuations in stock prices. Forecasting the stock market has been a widely discussed subject for many years, however, dependable predictions have been challenging to obtain. According to financial analysts, the spread of information to the public can potentially impact the stock price of a company. LI Bing et al. employed a five-category system consisting of Positive+, Positive, Neutral, Negative, and Negative- to categorize tweets. The categorization was determined by analyzing the structure of each tweet as a compilation of words and phrases. Subsequently, a chi-squared test was applied for association rule mining to identify potential correlations between public sentiment and stock market performance. The study collected around 15 million Twitter records that referenced any of the 30 designated organizations. The hourly fluctuations in stock prices were forecasted through the classification of tweets into distinct mood categories. LI Bing et al.'s method exhibited higher accuracy in comparison to SVM, C4.5, and Naive Bayes. While LI Bing et al.'s method showed promising results, one potential limitation could be the generalizability of their findings. The study collected data from a specific set of organizations and may not be representative of the overall stock market. Additionally, the use of such volume of Twitter data may not always be feasible, as the volume and sentiment of tweets can vary widely depending on current events or trends. Therefore, it is essential to consider the context and limitations of the data source when interpreting the results of sentiment analysis for stock market forecasting.

The remainder of this paper follows a structured format. In section three, we provide a detailed account of the dataset used for analysis and the data pretreatment techniques employed to prepare it for analysis. Section four presents a comprehensive outline of our general approach to addressing the research problem, with subsequent sections delving into specific aspects of the methodology in greater detail. Section five focuses on the techniques of sentiment analysis used in our study, while section six provides a detailed description of our proposed model. In section seven, we present the results of our inquiry, including the implementation results of our model. In section eight, we scrutinize and analyze the results, offering a discussion of their implications. The conclusion section provides a summary of our research findings and their scientific significance. Finally, in the future scope section, we discuss potential avenues for further research and opportunities to expand our understanding in this domain.

## IV. DATASET

This comprehensive project aims to explore the intricate relationship between social media sentiment, specifically on Twitter, and stock market performance for four leading companies: Apple (AAPL) and Tesla (TSLA). To achieve this, the Twitter API and Yahoo Finance API were utilized to compile various datasets, including stock market data files from Yahoo Finance, such as AAPL.csv, TSLA.csv, AMZN.csv, and GOOG.csv, and files containing relevant tweets, like tweets_apple.csv and tweets_tesla.csv.

In the realm of the stock market, companies are denoted by ticker symbols. As an example, the stock symbol for Apple Inc. is represented by the ticker 'AAPL'. The corpus of data that has been collected for this project comprises a significant amount of Twitter posts, exceeding 300,000 tweets for a sole company over a 12-month period. In order to identify the tweets with the highest potential to affect public opinion and stock valuations, they were categorized according to the number of retweets they received. This methodology enables prioritizing the tweets with higher significance, which have a greater probability of influencing the stock prices of the concerned companies.

Yahoo Finance is a reliable platform that provides financial information, statistics, and analysis. The platform includes a collection of stock quotes, press releases, financial reports, and proprietary content, rendering it a reliable resource for acquiring both past and present market data. The utilization of Yahoo Finance is essential in this study as it serves as a dependable source of stock market information for the analyzed corporations.

Snscrape is a Python library designed for web scraping tweets from Twitter. As a potent tool capable of extracting large quantities of data, it encompasses tweets, user details, and various metadata. In this project, snscrape has been utilized to collect tweets associated with the four companies being examined. These tweets serve as the basis for analyzing the potential influence of social media sentiment on stock market performance.

The primary aim of this project is to investigate the potential influence of social media sentiment, specifically on Twitter, on the stock market behavior of Apple and Tesla. The utilization of the Twitter API and Yahoo Finance API has facilitated the acquisition and analysis of datasets that aid in the examination of this correlation. The project's objective is to enhance impact assessment by prioritizing content with the highest number of retweets. As an analysis of the correlation between Twitter activity and the stock market performance of Apple and Tesla, the study has utilized the Twitter API and Yahoo Finance API to gather and analyze various datasets in order to examine the correlation between them. The project's objective is to conduct a comprehensive impact analysis on the most retweeted content. The study seeks to clarify the influence of social media sentiment on shaping public perception and its consequential effect on the stock market performance. The findings obtained from this study may hold significant worth for investors, financial experts, and corporations as they strive to comprehend and utilize the influence of social media in the constantly changing financial environment.

| Feature | Description |
|---|---|
| rawContent | The unfiltered text or message of a social media post. |
| retweetCount | The number of times a social media post has been shared by other users. |
| viewCount | The total number of times a video or content has been viewed by users. |
| hashtags | Keywords or phrases preceded by a pound sign (#) used to categorize content on social media platforms. |
| date | The specific day, month, and year when an event occurred or a piece of content was created or shared. |

Table 1 Tweet data description

| Stock Price | Description |
|---|---|
| Open_value | The initial trading price of the first stock |
| Close_value | The price of the last stock trade |
| Low_value | The lowest trading price of the stock |
| High_value | The highest trading price of the stock |

Table 2 Stock market data description

## V. METHODOLOGY

### A. Data Preparation and Modelling

In this research, our primary objective was to investigate the potential association between social media sentiment and stock market performance for selected companies, namely Apple and Tesla. We employed a systematic methodology that encompassed four key stages: data collection, preprocessing, sentiment analysis, and predictive modeling. The methodology was carefully designed to create a comprehensive dataset that combined both stock market data and social media sentiment, facilitating the development of a predictive model that could provide insights into the potential interplay between social media sentiment and stock market trends for the analyzed companies. The main stages of our methodology are detailed below:

Data Collection: We utilized the Twitter API and Yahoo Finance API to gather a comprehensive set of tweets related to the companies of interest and their corresponding stock market data. The data was collected over several days, capturing critical stock market aspects such as high, low, open, close, and volume values for each corporation.

Data Preprocessing: This stage involved cleaning and organizing the collected data to facilitate further analysis. Text preprocessing techniques, such as tokenization, stopword removal, and stemming, were applied to the tweets to prepare them for sentiment analysis. Additionally, the stock market data was processed to create new variables representing the previous day's values, enabling the analysis of potential relationships between past and current data. Namely, in the preprocessing stage, we structured the acquired data and introduced new columns in the dataset to signify the prior day's stock market variables, including "Prev Open," "Prev Close," among others. We then shifted both the stock market data and the consolidated tweets data by one day, effectively aligning the previous day's values and sentiment scores with the corresponding current day's stock market information.

Predictive Modeling: Utilizing the preprocessed dataset containing stock market values and sentiment scores, we devised an appropriate predictive model (e.g., regression, time series, or machine learning) to forecast the high, low, open, close, and volume metrics for the current day based on the input features. This model was engineered to capitalize on the insights derived from the previous day's tweets and stock market data to unveil valuable information concerning the relationship between social media sentiment and stock market patterns.

By following this systematic approach, we created a comprehensive dataset that integrated stock market data and social media sentiment, enabling the development of a predictive model that provided insights into the potential relationship between social media sentiment

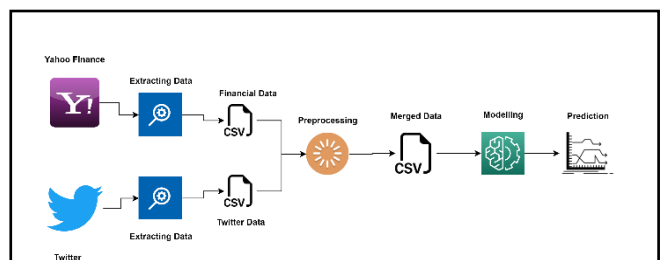

Figure 1 Methodology

### B. Sentiment Analysis

The present study involved the preprocessing and cleansing of Twitter data, followed by the application of sentiment analysis using the BERT model. BERT (Bidirectional Encoder Representations from Transformers) is a state-of-the-art natural language processing model that exhibits exceptional performance across various applications, such as sentiment analysis. In order to obtain sentiment scores for the tweets, we integrated the use of BERT within our data preprocessing pipeline using the following procedures:

Employing the BERT model for sentiment analysis, the function called get_sentiment_scores was designed to accept a given input string and generate a dictionary containing sentiment scores for positive, negative, and neutral emotions. This code leverages the BERT tokenizer and model to create inputs, compute logits, and apply the softmax function to yield probability scores for each sentiment category. The cleaned Twitter data was imported into a DataFrame using the pandas library. This dataset consists of dates, tweet summaries, and sentiment analyses. Adding a sentiment score column. New columns were introduced to the DataFrame to store the sentiment scores obtained from the BERT model. To track the public's reaction to each tweet, we divided the data into three distinct categories: negative, neutral, and positive. Iteratively assessing social media posts and calculating affective valence measures, we performed iteration on each entry in the DataFrame and extracted the text data from the column containing the tweets posted on the previous day. The unicodedata.normalize function was used to reformat the text in a way that is compatible with the BERT model. The sentiment of the text was determined by invoking the get_sentiment_scores function. After sentiment analysis, the DataFrame was updated by inserting the scores into their respective columns.

Saving the sentiment scores, once the sentiment scores were calculated and added to the DataFrame, we saved the updated DataFrame to a new CSV file. This file contains the preprocessed tweet data along with the sentiment scores obtained using the BERT model.

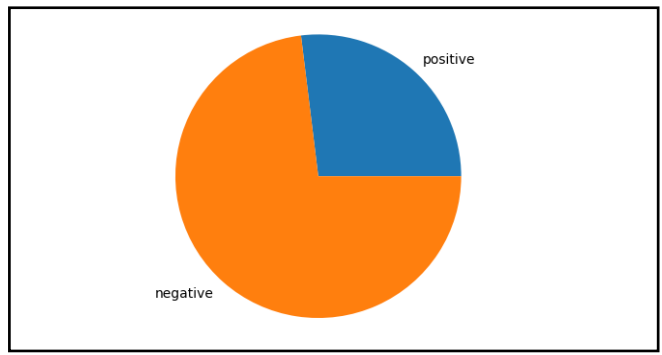

Figure 2 Positive Negative rate Tesla Tweets

By incorporating the BERT model for sentiment analysis, we were able to derive valuable insights from the tweet data and better understand the public sentiment towards the selected companies. The sentiment scores obtained through this process played a crucial role in our subsequent analysis and prediction modeling, as they provided an additional source of information to help determine the potential relationship between social media sentiment and stock market performance.

## VI. PROPOSED MODEL

Utilizing the discovered data, we have devised a scientific model that integrates Twitter sentiment metrics and historical stock price information to forecast the stock prices for the next trading day. By employing the Python Keras library, we have constructed a Long Short-Term Memory (LSTM) model. The initial step involved extracting relevant data from the 'tweets_stocks' DataFrame. The data was systematically organized into separate columns, which included the current price, open, high, low, close, volume, and sentiment values (negative, neutral, and positive) for the previous day.

First, we defined the independent and dependent variables for our model, followed by loading the data into a DataFrame named 'df'. The input variables consisted of the open, high, low, close, volume, and sentiment scores (negative, neutral, and positive) from the previous day. The target metric is represented by the closing price of the stock for the day. The data was preprocessed using the MinMaxScaler from the 'sklearn' library to normalize the input and target variables to a range between 0 and 1. After data cleaning and normalization, 80% of the entire dataset was allocated for use in the training process. Subsequently, the dataset was divided into two distinct subsets, one for training and the other for testing purposes.

In order to ensure compatibility with the LSTM model, it was crucial to transform the input data into a 3D format consisting of samples, time steps, and features [13]. Upon completion of data preparation, the LSTM model was established utilizing the Sequential API sourced from the Keras library. The architecture of the model consisted of three Long Short-Term Memory (LSTM) layers, with each layer comprising 100 units. In order to address the potential issueof overfitting, we integrated Dropout layers featuring a dropout rate of 0.2 amidst the LSTM layers. The final layer of the neural network was comprised of a singular Dense layer, possessing a solitary unit, which was assigned the responsibility of forecasting the closing stock price for the subsequent day. The loss function utilized in our study was the mean absolute error (MAE), and the Adam optimizer was employed for the model's training.

Subsequently, the LSTM model was trained on the provided training dataset. The training process of the model involved 100 epochs with a batch size of 4. The model's performance was evaluated by assessing the testing data as validation data. The value of the verbose parameter was set to 2, resulting in a concise representation of the training procedure.

The LSTM model generated is proficient in forecasting the subsequent day's stock price by incorporating both the preceding day's stock price data and the sentiment values obtained from Twitter. The aforementioned methodology facilitates a more all-encompassing comprehension of the variables that impact the oscillation of stock prices, thereby enhancing the efficacy of decision-making processes within financial markets. The study showcases the potential of augmenting the precision and efficacy of stock price prediction models by integrating diverse data sources and techniques, including sentiment analysis and deep learning methods like LSTM.

Figure 3 LSTM model summary

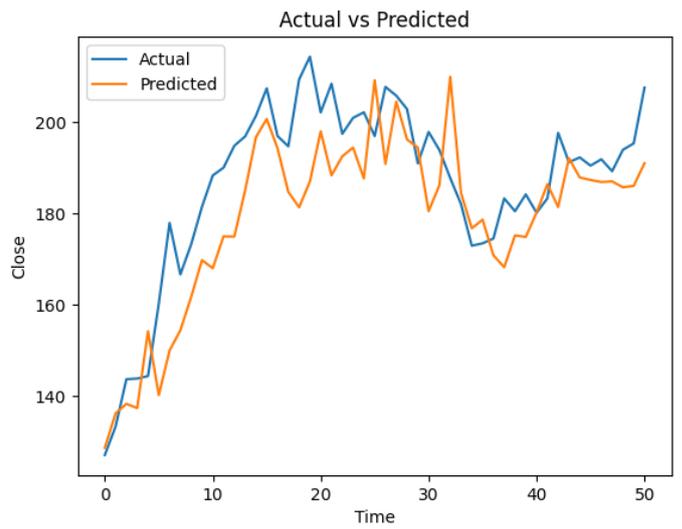

Figure 6 Actual vs Predicted (TESLA)

## VII. RESULTS

Figure 4 LSTM implementation (TESLA)

Figure 7 LSTM implementation (APPLE)

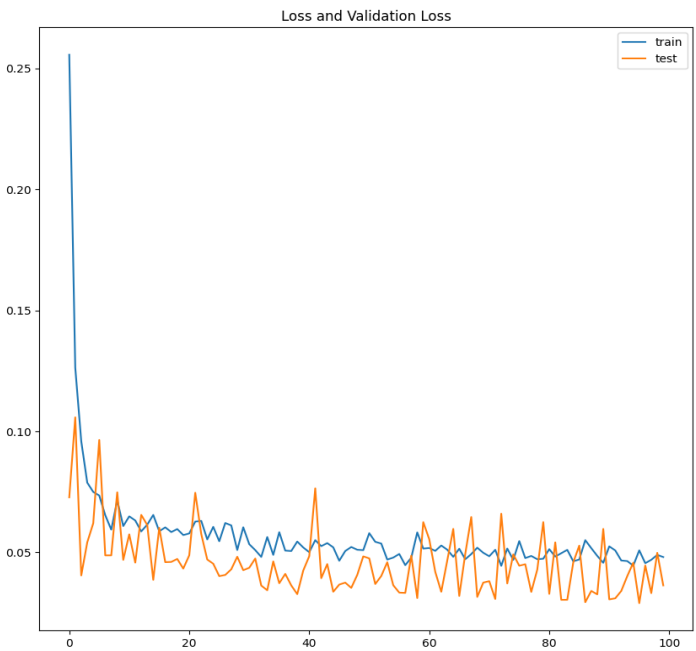

Figure 5 Train loss and test loss (TESLA)

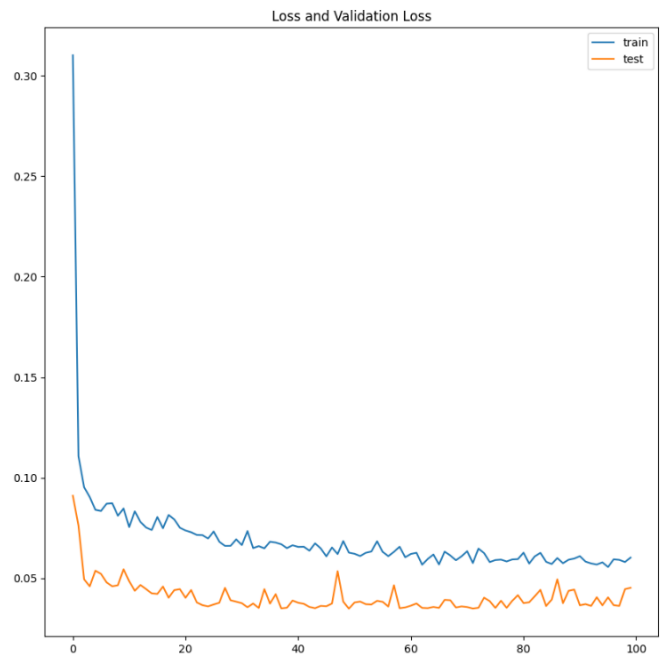

Figure 8 Train loss and test loss (APPLE)

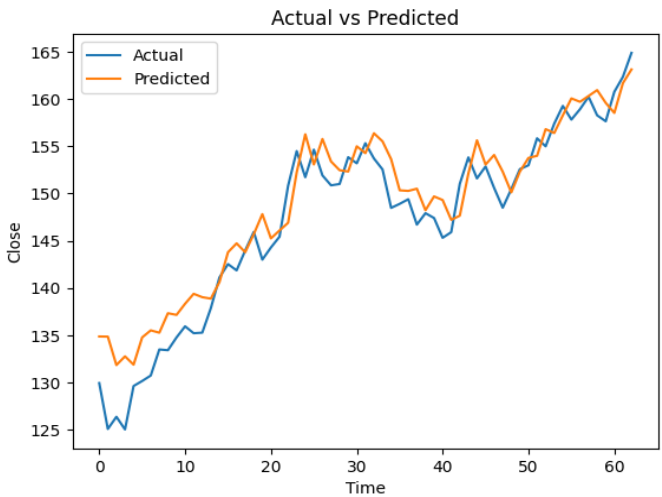

Figure 9 Actual vs Predicted (APPLE)

VIII. DISCUSSION OF RESULTS

The obtained results from our analysis provide valuable insights into the performance of the LSTM model when applied to the prediction of stock prices for Tesla and Apple based on historical stock data and sentiment scores derived from Twitter. In this section, we discuss the outcomes of the model training and testing process and highlight the implications of the findings for future research.

The training process for both Tesla and Apple conducted over 100 epochs. The model for Tesla achieved a final training loss of 0.0586 and a validation loss of 0.0403, whereas the model for Apple had a final training loss of 0.0480 and a validation loss of 0.0363. The lower validation losses indicate that the models have learned to make relatively accurate predictions on the unseen data. It is worth noting that the differences in the loss values between training and validation sets are relatively small, suggesting that the models have not overfitted the training data.

The mean absolute error (MAE) values obtained for Tesla and Apple were 9.93 and 2.47, respectively. These values represent the average absolute difference between the predicted and actual stock prices. The relatively low MAE values for both companies demonstrate the model's capability to make reasonably accurate predictions, considering the inherent volatility and complexity of the stock market.

The actual vs. predicted stock price plots for both Tesla and Apple further support the performance of the LSTM models. The plots show that the predicted stock prices follow the general trend of the actual stock prices, capturing the upward and downward movements over time. Although the model predictions may not be perfect, the overall trends observed in the plots suggest that the LSTM model can capture the complex relationships between historical stock prices, sentiment scores, and future stock prices.

The analysis is based on a small dataset, and adding more data from a wider time range could enhance the performance of the models. Additionally, by experimenting with various model architectures, hyperparameters, and feature engineering techniques, the models could be further optimized. It would also be intriguing to investigate the incorporation of additional external inputs that could enhance the model's prediction capabilities, such as macroeconomic data and news relevant to a particular industry.

Furthermore, the study focuses on two specific companies, Tesla and Apple, which are well-established technology firms. The performance of the LSTM models may differ when applied to other companies, industries, or market segments. Future research could extend the analysis to a broader range of companies and sectors, as well as investigate the applicability of the model to different financial instruments, such as options, futures, and cryptocurrencies.

In conclusion, our analysis demonstrates the potential of LSTM models to predict stock prices for Tesla and Apple using historical stock data and sentiment scores derived from Twitter. The relatively low MAE values and the consistency of the predicted trends with the actual stock prices suggest that the LSTM models can capture the complex dynamics underlying stock price movements. While there are limitations to the current study, the findings provide a foundation for future research on the application of deep learning techniques to financial market prediction and the integration of social media sentiment data into stock market analysis.

IX. CONCLUSION

In this study, we aimed to develop a predictive model for stock prices using historical data and sentiment scores derived from Twitter. We focused on four major companies, namely Apple and Tesla, and collected relevant datasets containing financial information and tweets associated with the companies. The primary objective of this research was to investigate the potential of combining traditional stock market data with sentiment analysis of social media data to improve stock price prediction accuracy.

The Yahoo Finance API and the Twitter API were used to get the information, and the ticker symbols were used to find the companies. We analyzed the data we collected by grouping tweets together based on how many times they were retweeted. This made sure that the most important tweets were included in the study. The BERT model, which is a powerful way to handle natural language, was then used to assign sentiment scores to the collected tweets. We chose to look at groups of tweets instead of individual tweets because we thought this would give us a better idea of how the Twitter public as a whole felt.

We looked at the challenge of forecasting stock prices as a multivariate time series forecasting problem, with the goal variable being dependent on its previous values as well as other variables such as sentiment scores. In order to accomplish this, we made use of a model of a neural network called a Long Short-Term Memory (LSTM) network, which is effective for dynamic time series forecasting and is able to properly deal with the issues of vanishing gradients that frequently impact recurrent neural networks.

The dataset was split into training and testing sets, with 80% of the data used for training and the remaining 20% for testing. The LSTM model was trained with various configurations to observe the impact of increasing LSTM blocks in a hidden layer and the effects of adding a Dropout layer. The model's performance was evaluated using the mean absolute error (MAE) metric, which measures the average magnitude of errors in the predictions.

The results obtained from our LSTM model demonstrated the model's ability to predict stock prices based on historical data and sentiment scores derived from Twitter. The MAE was found to be 9.93, indicating a moderate level of accuracy in the predicted stock prices. The visualization of the actual vs. predicted stock prices revealed that the model captured the general trends in the stock prices, though there were instances where the predicted values deviated from the actual values.

Several factors might have contributed to the observed discrepancies in the predictions. The relatively small size of the dataset used for training the model could limit its ability to learn the complex relationships between the input features and the target variable. The choice of input features might also impact the model's prediction accuracy, as stock prices are influenced by various factors beyond historical data and Twitter sentiment. Additionally, the choice of the model architecture and hyperparameters can significantly impact the model's performance, and alternative configurations might yield better results [19].

In conclusion, this study demonstrated the potential of combining traditional stock market data with sentiment analysis of social media data to predict stock prices. The LSTM model employed in this research captured the general trends in the stock prices, indicating its ability to predict stock price movements to some extent. However, there remains room for improvement through the expansion of the dataset, incorporation of additional input features, and exploration of alternative model architectures and hyperparameters.

This study adds to the expanding body of literature on the topic of predicting stock prices using machine learning and social media sentiment analysis. It aids stock market participants and investors in making better selections with the help of predictive models. The results also indicate that traditional financial models can benefit from embracing new data sources, such as social media sentiment, leading to improved stock price forecasting in the future.

## X. FUTURE SCOPE

The results of this study show promising potential for the use of sentiment analysis in combination with traditional stock market data for stock price prediction. However, there is ample scope for further research and improvements in this area, which can contribute to the development of more accurate and reliable predictive models. The following future research directions can be considered:

Expanding the dataset: The relatively small dataset used in this study might have limited the model's ability to learn the complex relationships between input features and the target variable. Future research could involve collecting more extensive datasets, spanning a longer time period and covering a larger number of companies across various sectors. This would help improve the model's performance by providing a more comprehensive understanding of the factors influencing stock prices.

Incorporation of additional input features: Stock prices are influenced by various factors beyond historical data and Twitter sentiment. Future research could explore the inclusion of other relevant input features, such as macroeconomic indicators (e.g., GDP growth, inflation rates), company-specific financial ratios (e.g., price-to-earnings ratio, debt-to-equity ratio), and market news sentiment, to capture a broader range of factors that may impact stock prices.

Exploration of alternative model architectures and hyperparameters: The choice of model architecture and hyperparameters can significantly impact the performance of the predictive model. Future studies could investigate alternative model architectures, such as convolutional neural networks (CNNs) or attention-based models like Transformers, which might offer better performance in certain contexts. Additionally, the use of advanced optimization techniques, such as Bayesian optimization [14] or genetic algorithms, for hyperparameter tuning could lead to improved model performance.

Comparison with other sentiment analysis techniques: This study employed the BERT model for sentiment analysis, which is a powerful natural language processing technique. However, there are other sentiment analysis approaches, such as using pre-trained models like VADER [15] or employing other NLP techniques like word embeddings (e.g., Word2Vec, GloVe), which could also be explored in future research for comparison purposes.

Investigating the impact of event-driven sentiment: Stock prices can be significantly influenced by specific events, such as product launches, earnings announcements, or regulatory changes. Future research could delve into the role of event-driven sentiment in stock price prediction, by identifying and incorporating event-related information from various sources, including news articles, press releases, and financial reports.

Ensemble learning and model stacking: Combining the predictions of multiple models can often lead to improved prediction performance. Future research could explore the use of ensemble learning techniques, such as bagging, boosting, or stacking, to combine the predictions from various models, including those based on traditional financial data, sentiment analysis, and other relevant features.

Real-time prediction and deployment: This study focused on predicting stock prices based on historical data and sentiment scores. Future research could explore the development of real-time prediction models, which could ingest live streaming data from sources such as Twitter and financial news feeds. Additionally, the deployment of such models in real-world applications, such as trading platforms or investment advisory tools, could be investigated.

Cross-market analysis: This study focused on stock price prediction in the context of individual companies. Future research could extend this approach to cross-market analysis [16], investigating the relationship between sentiment scores and stock price movements across different markets or sectors. This could provide valuable insights into how sentiment influences the dynamics of the broader financial market.

In summary, by addressing the limitations identified in this study and incorporating new techniques, data sources, and methodologies, future research can contribute significantly to the development of more accurate, reliable, and actionable stock price prediction models, benefiting investors, stakeholders, and the financial industry as a whole.

# XI. APPENDIX

## A. Data Preprocessing:

```python
def clean_tweet(tweet):
  if type(tweet) == np.float:
    return ""
  temp = tweet.lower()
  temp = re.sub("'", "", temp) # to avoid removing contractions in english
  temp = re.sub("@[A-Za-z0-9_]+","", temp)
  temp = re.sub("#[A-Za-z0-9_]+","", temp)
  temp = re.sub(r'http\S+', '', temp)
  temp = re.sub('[()!?]', ' ', temp)
  temp = re.sub('\[.*?\]',' ', temp)
  temp = re.sub("[^a-z0-9]"," ", temp)
  return temp
```

## B. Sentiment Analysis:

```python
# Function to get sentiment scores using BERT
def get_sentiment_scores(text):
    inputs = tokenizer.encode_plus(text, return_tensors='pt', truncation=True, padding=True)
    outputs = model(**inputs)
    scores = torch.softmax(outputs.logits, dim=1).tolist()[0]
    return {
        'negative': scores[0],
        'neutral':  scores[1],
        'positive': scores[2]
    }
# Load the tweet data into a dataframe
ccdata = cdata_grouped.copy()
# Add new columns for the sentiment scores
ccdata['negative'] = ''
ccdata['neutral'] = ''
ccdata['positive'] = ''

# Iterate over each tweet and get the sentiment scores
for index, row in ccdata.iterrows():
    try:
        sentence = unicodedata.normalize('NFKD', row['Previous Day Tweets'])
        sentiment_scores = get_sentiment_scores(sentence)
        ccdata.at[index, 'negative'] = sentiment_scores['negative']
        ccdata.at[index, 'neutral'] = sentiment_scores['neutral']
        ccdata.at[index, 'positive'] = sentiment_scores['positive']
    except TypeError:
        print(f"Error processing tweet at index {index}: {row['Previous Day Tweets']}")
```

## C. Predictive Modeling:

```python
# define the input and target variables
X = df[['Prev Open', 'Prev High', 'Prev Low', 'Prev Close', 'Prev Volume',
        'negative', 'neutral', 'positive']].values

y = df['Close'].values

# normalize the data using MinMaxScaler
scaler = MinMaxScaler(feature_range=(0, 1))
X = scaler.fit_transform(X)
y = scaler.fit_transform(y.reshape(-1, 1))

# define the size of the training set
train_size = int(len(df) * 0.8)

# split the data into training and testing sets
X_train, X_test = X[:train_size, :], X[train_size:, :]
y_train, y_test = y[:train_size], y[train_size:]

# reshape the input data into 3D format (samples, time steps, features)
X_train = X_train.reshape((X_train.shape[0], 1, X_train.shape[1]))
X_test = X_test.reshape((X_test.shape[0], 1, X_test.shape[1]))

# define the LSTM model
model = Sequential()
model.add(LSTM(100, input_shape=(X_train.shape[1], X_train.shape[2]), return_sequences=True))
model.add(Dropout(0.2))
model.add(LSTM(100, return_sequences=True))
model.add(Dropout(0.2))
model.add(LSTM(100))
model.add(Dropout(0.2))
model.add(Dense(1))
model.compile(loss='mae', optimizer='adam')

# train the model
model_his = model.fit(X_train, y_train, epochs=100, batch_size=4, validation_data=(X_test, y_test), verbose=2
```